\documentclass[10pt,superscriptaddress,twocolumn,amsmath,amssymb,aps,prl]{revtex4}
\usepackage{mathrsfs}
\usepackage{graphicx}% Include figure files
\usepackage{dcolumn}% Align table columns on decimal point
\usepackage{bm}% bold math
\usepackage{color}

\newcommand{\fig}[2]{\includegraphics[width=#1]{#2}}

\newcommand{\be}{\begin{equation}}
\newcommand{\ee}{\end{equation}}
\newcommand{\bea}{\begin{eqnarray}}
\newcommand{\eea}{\end{eqnarray}}

\begin{document}

\title{   Cobalt-dimer Nitrides - a Potential Novel Family of High Temperature Superconductors}

\author{Yuhao Gu}
\affiliation{Beijing National Laboratory for Condensed Matter Physics,
and Institute of Physics, Chinese Academy of Sciences, Beijing 100190, China}

\author{Kun Jiang}
\affiliation{Beijing National Laboratory for Condensed Matter Physics,
and Institute of Physics, Chinese Academy of Sciences, Beijing 100190, China}
\affiliation{School of Physical Sciences, University of Chinese Academy of Sciences, Beijing 100190, China}

\author{Xianxian Wu}
\affiliation{ Institute of Theoretical Physics, Chinese Academy of Sciences, Beijing 100190, China}
 
\author{Jiangping Hu}\email{jphu@iphy.ac.cn}
\affiliation{Beijing National Laboratory for Condensed Matter Physics,
and Institute of Physics, Chinese Academy of Sciences, Beijing 100190, China}
\affiliation{Kavli Institute of Theoretical Sciences, University of Chinese Academy of Sciences,
Beijing, 100190, China}

\date{\today}

\begin{abstract}
We  predict that the square lattice layer formed by [Co$_2$N$_2$]$^{2-}$ diamond-like units can host high temperature superconductivity. The layer appears in the stable ternary cobalt nitride, BaCo$_2$N$_2$.  The electronic physics of the material stems from Co$_2$N$_2$ layers where the dimerized Co pairs form a square lattice. 
The low energy physics near Fermi energy can be described by  an effective two  orbital model. Without considering interlayer couplings, the two orbitals are effectively decoupled. This electronic structure satisfies the ``gene" character  proposed for unconventional high temperature superconductors. We  predict that the leading superconducting pairing instability is driven from an extended $s$-wave (s$^\pm$) to a $d$-wave by hole doping, for example in Ba$_{1-x}$K$_x$Co$_2$N$_2$.  %In the transitional region, the material can host a mixed superconducting state such as d+is.  
This study provides a new platform to  establish the  superconducting mechanism of unconventional high temperature superconductivity. 
\\
\\
%{\color{red}\text{Keywords: high temperature superconductors, cobalt-dimers, nitrides, s(d)-wave pairing symmetry}}
\end{abstract}

\maketitle

A roadmap to search for or design unconventional high temperature superconductors (T$_c$s) has been recently proposed by us\cite{jpHuprx, jpHusciencebult}.  It is based on the idea that unconventional high T$_c$s  must host  a certain electronic environment  in which those $d$-orbitals of  transition metal atoms with the strongest in-plane coupling to the $p$-orbitals of anions  are isolated near Fermi energy.  Both cuprates\cite{Cu} and  iron-based superconductors\cite{Fe}   host such an environment.  This simple feature not only  guarantees the superexchange antiferromagnetic interactions that are generated through $d-p$ couplings can be maximized to provide superconducting pairing, but also explains why high T$_c$s are rare in nature.   Thus,  this electronic feature can be considered as the gene of  unconventional high T$_c$s. Following this understanding, a few cases for Cobalt or Nickel compounds have been theoretically proposed\cite{jpHuprx,Hu-Co2017,Le-BaCoSO-2017}. Unfortunately, none of these cases has been  realized experimentally because  the ``gene" condition is typically not  favored energetically. 

Isolating those $d$-orbitals to satisfy the ``gene" condition always costs significant energy.   Therefore, creating such an environment requires a specific lattice structure in which the energy cost  can be compensated.   For example, in cuprates, the energy cost of isolating  the d$_{x^2-y^2}$ e$_g$ orbital is compensated by lowering the 3t$_{2g}$  in the perovskite-type of structure.    In iron-based superconductors, the energy cost of isolating the t$_{2g}$ orbitals in a tetrahedron environment is reduced by  the hybridization of  the  e$_g$  and  some part of  t$_{2g}$ orbitals between two edge shared tetrahedrons.  This hybridization is critical in creating the ``gene" condition for  a partially filled $d$-shell.  The hybridization creates an empty   hybridized band at high energy while leaving two near half-filled bands attributed to pure  t$_{2g}$ orbitals  near Fermi energy in the d$^6$ configuration of Fe$^{2+}$. We have proposed that the ``gene" condition can be created in a layer formed by corner-shared tetrahedrons\cite{Hu-Co2017,hu2018diamond}, in which all three t$_{2g}$ can be isolated near Fermi energy for the d$^7$ configuration of Co$^{2+}$.  However, forming such a layer  is difficult because of too much energy cost in isolating all three anti-bonding t$_{2g}$ orbitals.

\begin{figure}
	\begin{center}
		\fig{3.4in}{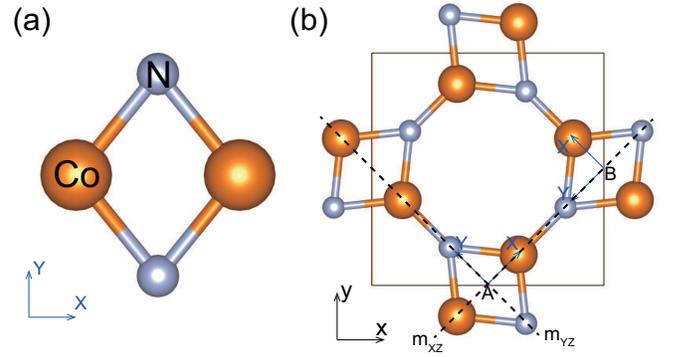}
		\caption{(a) Structure of Co$_2$N$_2$ dimer. (b) Structure of Co$_2$N$_2$ layer. The dashed black lines show the mirror symmetries. $x$-$y$ is the global coordinate in the conventional crystal structure and $X$-$Y$ coordinates are defined according to the direction of  the Co-Co bond.
			\label{Co2N2}}
	\end{center}
	%	\vskip-0.5cm
\end{figure}

  In this paper, we predict that a new ``gene" exists for  the d$^7$ configuration of Co$^{2+}$ in layers formed by [Co$_2$N$_2$]$^{2-}$ diamond-like units as shown in Fig.\ref{Co2N2}(a). The layers appear in a ternary cobalt nitride, BaCo$_2$N$_2$ (BCN), which  is structurally stable with minimum formation energy.  Cobalt atoms are dimerized in a  [Co$_2$N$_2$]$^{2-}$  unit.  The dimers form a two dimensional square lattice and their alignments alternate in two sublattices as shown in Fig.\ref{Co2N2}(b). We show that the electronic physics near Fermi energy can be described by  an effective two orbital model in the square lattice and satisfies the ``gene"  condition.  There are strong antiferromagnetic interactions when  electron-electron correlations are included.  In particular, with hole doping, the Fermi surfaces can be driven from  the iron-pnictide type with the simultaneous appearance of the hole and electron pockets\cite{Poltavets2009,Seo2008,Hirschfeld2011}  to the cuprate type with a single Fermi surface near half filling\cite{damascelli2003angle}. We predict that both extended $s$-wave($s^\pm$)\cite{Seo2008,Hirschfeld2011} and $d$-wave superconducting states can be formed in different doping regions.  The transitional region between these two states may  be featured  by a mixed $s$-wave and $d$-wave superconducting state. 
  
\begin{figure}
	\begin{center}
		\fig{3.4in}{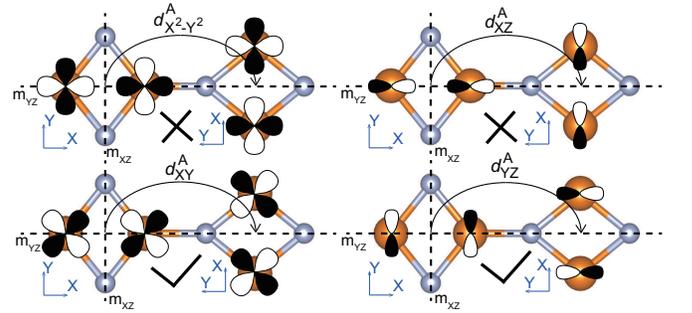}
		\caption{Sketch of intra-orbital NN hopping between anti-bonding molecular orbitals of Co$_2$N$_2$ dimers. The intra-anti-bonding-orbital hoppings between the $d_{X^2-Y^2}^A$/$d_{XZ}^A$ molecular orbitals of two NN dimers are forbidden due to the opposite mirror symmetry eigenvalues while those between the $d_{XY}^A$/$d_{YZ}^A$ molecular orbitals are allowed.
			\label{NNhop}}
	\end{center}
	%	\vskip-0.5cm
\end{figure}

 Before presenting the detailed calculation, we use the symmetry analysis to argue why the electronic physics in the single Co$_2$N$_2$ layer is described by a two orbital model and satisfies the ``gene" condition.
  In a  [Co$_2$N$_2$]$^{2-}$  unit in Fig.\ref{Co2N2}(a), the states from the $d$-orbitals form bonding (d$^B$) and anti-bonding (d$^{A}$) states.  It is obvious that the bonding states, as well as the states from the d$_{Z^2}$ orbital, have much lower energy and are irrelevant.   We can focus on the four d$^{A}$ states and classify them with respect to the XZ,YZ and XY mirror plane symmetries as shown in Fig.\ref{NNhop}. The states can be labeled by the symmetry characters as (d$_{XY}^{A}$$(-,-,+)$, d$_{YZ}^{A}$$(-,-,-)$, d$_{X^2-Y^2}^{A}$$(+,-,+)$, d$_{XZ}^{A}$$(+,-,-)$), where the signs represent the three mirror symmetries characters. In the first order approximation, we only need to consider the hoppings between the two nearest neighbor (NN) sites to determine the kinematics of these states. It is important to notice that the two dimers between two NN sites align perpendicularly. Therefore, only the electrons in those states with the same symmetry characters with respect to the XZ and YZ planes are allowed to hop between the two NN sites, namely, only the electrons in d$_{XY}^{A}$$(-,-,+)$ and d$_{YZ}^{A}$$(-,-,-)$
  can generate NN hoppings, as illustrated in Fig.\ref{NNhop}.  The intra-orbital NN hopping within the other two states is forbidden due to the perpendicular alignment of the two NN dimers.  We also notice  that the d$_{XY}$ and d$_{YZ}$  orbitals are those orbitals with stronger $d-p$ couplings. For the $d^7$ configuration of Co$^{2+}$, there are roughly two electrons left to fill the two bands.  This rough analysis suggests that  the electronic band structure of the Co$_2$N$_2$ layer  can be described by  a model with two decoupled orbitals and can satisfy the ``gene" condition.

\begin{figure*}
	\begin{center}
		\fig{7in}{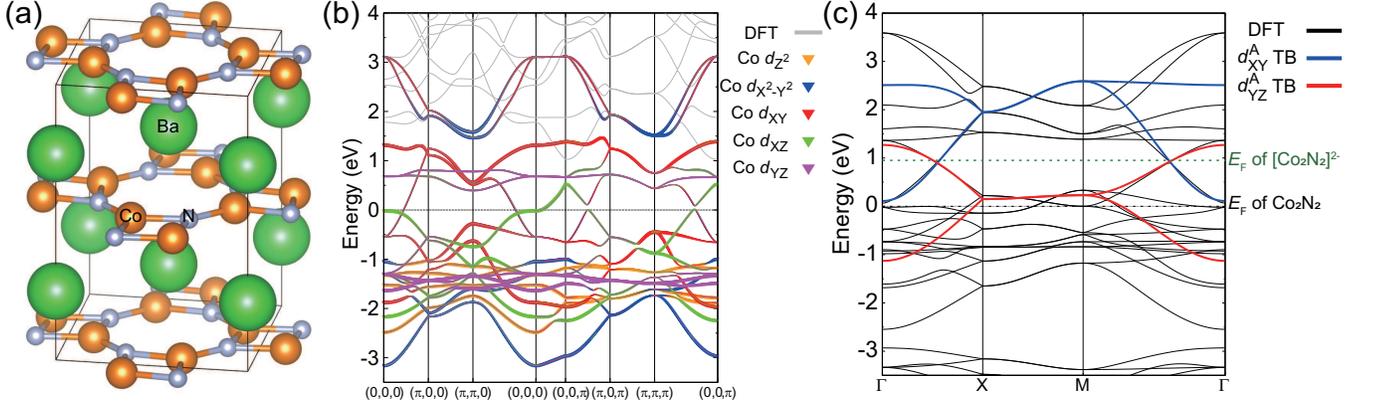}
		\caption{(a) Crystal structure of BaCo$_2$N$_2$ in the conventional cell. (b) DFT-calculated band structure (gray line) and projected weight of $d$-orbital-like WFs in its symmetric local coordinate ($X$-$Y$ coordinate). The coordinates in the abscissa are corresponding to the high symmetric $\textbf{k}$-points of the tetragonal conventional cell's Brillouin zone. (c) Band structures of Co$_2$N$_2$ monolayer from DFT and our effective TB model. The black dotted line represents the Fermi level ($E_{\rm F}$) of Co$_2$N$_2$ monolayer while the green dotted line represents the Fermi level ($E_{\rm F}$) of [Co$_2$N$_2$]$^{2-}$. 
			\label{BCN}}
	\end{center}
	%	\vskip-0.5cm
\end{figure*}

%{\it  ab-initio band structure } 
Now we perform the detailed DFT calculation of the electronic structure of the system with VASP\cite{Kresse1996,Kresse1999} and Wannier90\cite{mostofi2008wannier90,Marzari2012}. The Perdew-Burke-Ernzerhof (PBE)\cite{perdew1996} exchange-correlation functional is used in our calculations. The crystal structure of BaCo$_2$N$_2$ is displayed in Fig.\ref{BCN}(a) and the alkaline-earth metals are intercalated between Co$_2$N$_2$ layers, similar to iron pnictides. The Co$_2$N$_2$ layer has a planar structure, distinct from the  trilayer structure. One prominent feature is that the short distance between the nearest-neighbor (NN) Co sites induces a strong hybridization between them, generating bonding and anti-bonding states. Fig.\ref{BCN}(b) shows the orbital-resolved band structure from DFT calculations. The N $p$ orbitals, absent in the figure, are located around 3.5 eV below the Fermi level, while Co 3$d$ orbitals dominate from -3 eV to 3 eV. Despite moderate inter-orbital coupling, low-energy bonding states and high-energy anti-bonding states for each Co 3$d$ orbital can be identified. We also notice that  between -0.5eV to 0.5eV near the Fermi level, the electronic structure is simply represented by two bands with large dispersion, which is consistent with our previous qualitative analysis from symmetry. 

\begin{figure}
	\begin{center}
		\fig{3.4in}{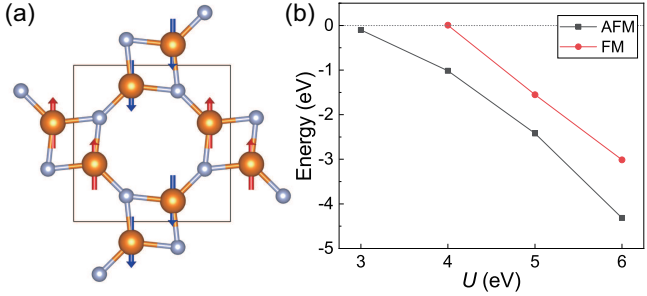}
		\caption{(a) The in-plane AFM order of Co$_2$N$_2$ layer. (b) DFT+$U$-calculated energies of different magnetic orders of BaCo$_2$N$_2$ relative to the paramagnetic state as a function of the onsite $U$. 
			\label{mag}}
	\end{center}
	%	\vskip-0.5cm
\end{figure}

%{\it two-orbital tight-binding model}
To obtain the effective model in the monolayer Co$_2$N$_2$, we first analyze its symmetry. The point group at A/B site is $D_{2h}$ and the anti-bonding states of Co $d_{YZ}$ and $d_{XY}$ belong to $A_u$ and $B_{1g}$ irreducible representations, respectively. They are decoupled in the monolayer due to the mirror reflection with respect to XY plane. Remarkably, there is a glide mirror operation $\{M_{xz/yz}|\frac{1}{2},\frac{1}{2}\}$, connecting the Co sites around A and B. This nonsymmorphic symmetry can be used to unfold the band structure into a one-sublattice lattice, analogous to iron-based superconductors. The effective Hamiltonian can be simplified into a two-orbital model on a square lattice. Based on the anti-bonding Co d$_{XY}$ and d$_{YZ}$ orbitals, the two-orbital tight-binding (TB) Hamiltonian reads,
\begin{eqnarray}
H_{t}&=&\sum_{\bm{k}\alpha\beta\sigma}\varepsilon_k^{\alpha\beta}
d_{\bm{k}\alpha\sigma}^\dagger d_{\bm{k}\beta\sigma} + \sum_{\bm{k}\sigma} e_\alpha d_{\bm{k}\alpha\sigma}^\dagger d_{\bm{k}\alpha\sigma},
\label{ht}
\end{eqnarray}
where the operator $d_{\bm{k}\alpha\sigma}^\dagger$ ($d_{\bm{k}\alpha\sigma}$) creates (annihilates) an electron with momentum $\bm{k}$ of orbital $\alpha$ and spin $\sigma$ and $e_\alpha$ is the onsite energy for the $\alpha$ orbital.
The hopping terms in Eq.~(\ref{ht}) are given by
\begin{eqnarray}
\varepsilon_k^{\alpha\beta}&=&(2t_\alpha (\cos k_X+\cos k_Y)+4t_\alpha^\prime \cos k_X \cos k_Y)\delta_{\alpha\beta}. \nonumber
\label{epsilons}
\end{eqnarray}
We fit our model to the DFT band structure of the monolayer Co$_2$N$_2$ and  the fitting parameters are provided in Table \ref{hopping}. The corresponding TB band structure is shown in Fig.\ref{BCN}(d), in comparison to the DFT band structure. A reasonable agreement is reached between them.

\begin{table}[t]
	\caption{Hopping parameters of the TB model in eV.}
	
	\begin{tabular}{c|c|c|c}
		% after \\: \hline or \cline{col1-col2} \cline{col3-col4} ...
		\hline
		\hline
		hopping integral & 1rd ($t_\alpha$) & 2nd ($t^\prime_\alpha$) & e$_\alpha$ \\	\hline
		intra-orbital $t_{XY}$ & -0.30 & -0.16 & 2.0 \\	\hline
		intra-orbital $t_{YZ}$ & -0.30 & -0.02 & 0.2 \\	\hline
	\end{tabular}
	\label{hopping}
\end{table}

%{\it Antiferromagnetic interactions}
As unconventional superconductivity is usually in the vicinity of magnetic orders, we further study  magnetic interactions in BCN by performing calculations including an onsite interaction $U$ in Co sites (DFT+$U$)\cite{liechtenstein1995density}.  
We find that there are strong antiferromagnetic (AFM) interactions between two NN dimers in the tetragonal lattice in the presence of electron-electron correlations. In Fig.\ref{mag}(a), we show the AFM configuration of the states and in Fig.\ref{mag}(b), we plot the energy gain of the state as a function of $U$ relative to the paramagnetic state. When $U\ge3$ eV, the AFM state starts to have energy gain and becomes the ground state. By mapping the energies of different magnetic configurations into the Heisenberg model, we can estimate that the AFM interaction $J$ between two NN dimers is about 16.0/13.5/20.4 meV when $U=4/5/6$ eV. This $J$ between NN dimers is similar to $J_2$ between second-nearest neighboring (SNN) Fe atoms in iron-based superconductors\cite{Lu2009,zjf2017}. This AFM interaction is clearly mediated by N atoms and arises from the superexchange mechanism.

%{\it t-j model}
 Then, we can apply a strong correlation method to uncover the superconducting property of  BCN material.
The correlation part of the Hamiltonian $H_I$ follows the two-orbital Hubbard model  \cite{castellani,georges,kk},
\begin{eqnarray}
	H_I&=U\sum_{i,\alpha}\hat{n}_{i\alpha\uparrow}\hat{n}_{i\alpha\downarrow}
	+\left(U'-{1\over 2}J_H\right)\sum_{i,\alpha<\beta}\hat{n}_{i\alpha}\hat{n}_{i\beta}
	\label{hi} \\
	&-J_H\sum_{i,\alpha\neq\beta}{\bf S}_{i\alpha}\cdot {\bf S}_{i\beta}
	+J_H\sum_{i,\alpha\neq\beta}d^\dagger_{i\alpha\uparrow}
	d^\dagger_{i\alpha\downarrow}d_{i\beta\downarrow}d_{i\beta\uparrow},
	\nonumber
\end{eqnarray}
where the intra and interorbital repulsion $U$ and $U^\prime$ are related by Hund's rule coupling $J_H$
through $U=U'+2J_H$. The correlation parameters are set as $U=2$eV and $J_H=0.1U$.

The strong correlation effects are studied using the multiorbital Gutzwiller projection method \cite{gebhard98,lechermann,sen}: $H=H_t+H_I\to H_G=P_GH_tP_G$, where $P_G$ is the finite-$U$ Gutzwiller projection operator that reduces the statistical weight of the Fock states with multiple occupations.
The projection can be conveniently implemented using the Gutzwiller approximation (GW)
\cite{gebhard98,lechermann}, which has been used in the multiorbital cobaltates \cite{cobaltate}, Fe-pnictides \cite{sen}, and the monolayer CuO$_2$ grown on Bi$_2$Sr$_2$CaCu$_2$O$_{8+\delta}$ substrate \cite{jiang}. In this approach, the Gutzwiller projected Hamiltonian is expressed as
\begin{equation}
	H_{G}=\sum_{k\alpha\beta\sigma}g_{\alpha\beta}^\sigma\varepsilon_k^{\alpha\beta}
	d_{k\alpha\sigma}^\dagger d_{k\beta\sigma} + \sum_{k\sigma}(e_\alpha +\lambda_\alpha^\sigma)d_{k\alpha\sigma}^\dagger d_{k\alpha\sigma}.
	\label{ght}
\end{equation}
The strong correlation effects are described by the orbital dependent hopping renormalization $g_{\alpha\beta}^\sigma$ and the renormalized crystal field $\lambda_\alpha^\sigma$, which must be determined self-consistently for a given electron density $n$.

The superexchange interactions involving spin-orbital  can also be derived and the Kugel-Khomskii type terms are written as
\begin{eqnarray}
	H_{\rm J-K}=\sum_{\langle ij\rangle}\biggl[ J{\bf S}_{i}\cdot{\bf S}_j
	&+&\sum_{\mu\nu} I_{\mu\nu} T_i^\mu T_j^\nu
	\label{hs} \\
	&+&\sum_{\mu\nu} K_{\mu\nu}({\bf S}_{i}\cdot{\bf S}_j)(T_i^\mu T_j^\nu)\biggr]
	\nonumber
\end{eqnarray}
where the $J$-term is the SU(2) invariant Heisenberg spin exchange coupling, while the terms proportional $I_{\mu\nu}$ and $K_{\mu\nu}$
describe the anisotropic orbital and spin-orbital entangled superexchange interactions respectively, since the orbital rotation symmetry is broken by the lattice in the hopping Hamiltonian $H_t$.
Decoupling the superexchange interactions by including all spin-singlet pairing order parameters,
\begin{equation}
	\Delta_{ij}^{\alpha\beta\dagger}=d_{i\alpha\uparrow}^\dagger d_{j\beta\downarrow}^\dagger-d_{i\alpha\downarrow}^\dagger  d_{j\beta\uparrow}^\dagger,
	\label{pairingfield}
\end{equation}
we can arrive at the effective Hamiltonian as
\begin{eqnarray}
	H=P_GH_tP_G&-&\sum_{\langle ij\rangle}\biggl[{J_s\over2}\sum_{\alpha\beta}\Delta_{ij}^{\alpha\beta\dagger} \Delta_{ij}^{\alpha\beta}
	\nonumber  \\
	&+&{K\over2}\sum_{\alpha\neq\beta}(\Delta_{ij}^{\alpha\alpha\dagger}\Delta_{ij}^{\beta\beta}+\Delta_{ij}^{\alpha\beta\dagger}
	\Delta_{ij}^{\beta\alpha})\biggr]
	\label{htjk} \\
	%&-&{K_s\over2} \sum_{\alpha\neq\beta}\eta_{ij} \Delta_{ij}^{\alpha\alpha\dagger}(\Delta_{ij}^{\alpha\beta}+\Delta_{ij}^{\beta\alpha})+h.c.\biggr],
	\nonumber
\end{eqnarray}

\begin{figure}
	\begin{center}
		\fig{3.4in}{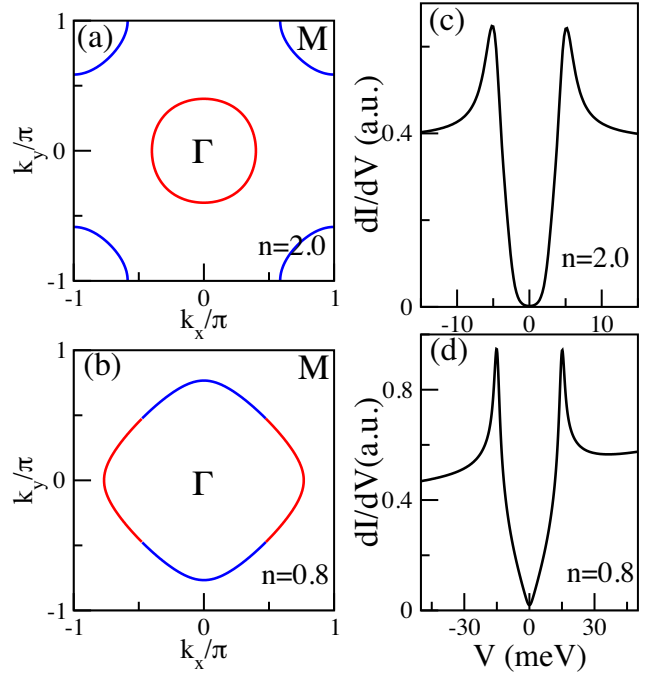}
		\caption{(a) Fermi surface of BCN at $n$=2.0. An s$^\pm$-wave SC is obtained here. The Fermi pocket with red color corresponds to the $\Delta(k)>0$ while the Fermi pocket with blue color corresponds to the $\Delta(k)<0$  patch.
			 (b) Fermi surface of BCN at $n$=0.8. A $d$-wave SC is obtained here. The FS with red color corresponds to the $\Delta(k)>0$ patch while the FS with blue color corresponds to the $\Delta(k)<0$  patch.
			 (c) Local tunneling density of states at $n$=2.0 with gap $\Delta=5.2$ meV. A thermal broadening of 0.5meV is used. 
			 (d)  Local tunneling density of states at $n$=0.8 with gap $\Delta= 15.2$ meV. 
			\label{fig3}}
	\end{center}
	%	\vskip-0.5cm
\end{figure}

Using the GW approximation, we found two different regimes separating by a Lifshitz transition when hole doping the BCN. This Lifshitz transition point $n_L$ is shifted from $0.86$ at $U=0$ and $1.12$ at $U=2$ eV owing to the correlation-induced crystal field renormalization.
When electron density $n$ is great than $n_L$, there are one hole pocket centering around the $\Gamma$ point and one electron pocket centering around the $M$ point respectively, as shown in Fig.\ref{fig3}(a) at $n$=2.0.
On the other hand, when $n$ is less than $n_L$, a Fermi surface (FS) centering around the $\Gamma$ point is arrived, as shown in Fig.\ref{fig3}(b) at $n$=0.8. These two regimes have different superconducting ground states.

In the low doping region, the physical property of BCN is similar to the iron-based SC with two electron pockets. In iron-based SC, a nodaless $s^{\pm}$ wave SC is obtained from both strong coupling and weak coupling studies\cite{Seo2008,Hirschfeld2011}.
By choosing $J_s=200$meV and $K=80$meV, a  $s^{\pm}$ wave with extend $s$-wave factor $\cos(k_x)+\cos(k_y)$ is also achieved from above self-consistent mean field study. As shown in Fig.\ref{fig3}(a), the pairing gap $\Delta(k)$ at Fermi pocket around $\Gamma$ is positive labeled with red color while  $\Delta(k)$ at Fermi pocket around $M$ is negative with blue color. To compare with scanning tunneling microscope, we also calculate the tunneling density of states (DOS). As shown in Fig.\ref{fig3}(c), a U-shape DOS spectrum is obtained with a coherent peak at $\Delta=5.2$ meV, which further confirms the nodaless nature of this regime.

In the above region, both bands are still away from the half-filling of each band. Hence, the superconducting gap is relatively weak. After the Lifshitz transition $n_L$, the lower band begins to close to its half-filling. Since the upper band is empty, the two-orbital model is downfolded to the single orbital model. In this regime ($n<n_L$), a physical property similar to the cuprates SC is achieved. We choose the filling at $n$=0.8, which is close to the optimal doping of cuprates. In this case, a  $d$-wave superconductor with form factor $\cos(k_x)-\cos(k_y)$ is obtained as shown in Fig.\ref{fig3}(b). The  $\Delta(k)$ around the FS breaks into four different patches, where the red patches show a positive gap and the blue patches show a negative gap.
Furthermore, a V-shape tunneling DOS is calculated in Fig.\ref{fig3}(d), which is a hallmark for $d$-wave SC. A much larger SC gap with $\Delta= 15.2$ meV  is obtained. Hence, a high-temperature SC with $d$-wave pairing can be achieved in the large hole doping BCN material. 
In the transitional region, a mixed $s$-wave and $d$-wave superconducting state may also emerge owing to the high-order coupling between $s$-wave and $d$-wave in the free energy \cite{sd1,sd2,sd3}.

In summary, we proposed a new family of high temperature superconductors consisting of [Co$_2$N$_2$]$^{2-}$ diamond-like units.
These [Co$_2$N$_2$]$^{2-}$ units from a square lattice layer. The low energy physics near Fermi energy of the layer is described by an effective decoupled two orbital model.  This electronic structure satisfies the ``gene" character  proposed for unconventional high temperature superconductors as both orbitals isolated near Fermi energy  generate large dispersions through $d-p$ couplings. 

The system can bridge the gap between the electronic physics of iron-based SC and cuprates. Without doping ($n$=2.0), the Fermi surface topology of the system is very similar to those of iron-pnictides with one electron pocket at the Brillouin zone center and one hole pocket at the Brillouin zone corner. In this case, an extended $s$-wave (s$^\pm$) pairing similar to iron-based SC is found using a strong-coupling mean-field study. However, at large hole doping close to $n$=1.0, $d$-wave superconductivity similar to cuprates is achieved.  The transitional region between these two SC states will be an extremely interesting region to study both theoretically and experimentally in the future. 

The layer is stabilized in BaCo$_2$N$_2$.  BaCo$_2$N$_2$ is the stablest Ba-Co-N ternary structure in its stoichiometric ratio according to the Materials Project database \cite{Materials_Project}. The hole doping can be achieved by substituting Ba with K as Ba$_{1-x}$K$_x$Co$_2$N$_2$. Thus, successfully synthesizing BaCo$_2$N$_2$ or related materials can establish a platform to decode the superconducting mechanism of unconventional high temperature superconductivity.

%{\it Acknowledgements}
This work is supported by the National Key Basic Research Program of China  (Grant
No. 2017YFA0303100), National Science Foundation of China (Grant No. NSFC-11888101, No. NSFC-12174428),  the Strategic Priority Research Program of Chinese Academy of Sciences (Grant
No. XDB28000000), and the Chinese Academy of Sciences through the Youth Innovation Promotion Association.

\end{document}